\documentclass[pra,twocolumn,aps,amssymb,footinbib]{revtex4}
\usepackage{amssymb}

\usepackage{amssymb}

\usepackage{graphicx}
\usepackage{amsmath}
\usepackage{times}
\usepackage{color}

\definecolor{lightblue}{rgb}{0,0.5,0.8}
\definecolor{darkgreen}{rgb}{0,0.5,0}

\begin{document}

\title{ Bragg spectroscopy of trapped one dimensional strongly interacting bosons in  optical lattices:
 Probing the cake-structure }

\author{ Guido Pupillo$^{1}$, Ana Maria Rey$^{2}$ and Ghassan George Batrouni$^{3}$}

 \affiliation{$^{1}$Institute
for Quantum Optics and Quantum Information of the Austrian Academy
of Sciences, 6020 Innsbruck, Austria}

 \affiliation{$^{2}$Institute for Theoretical
Atomic, Molecular and Optical Physics, Harvard-Smithsonian Center of
Astrophysics, Cambridge, MA, 02138, USA}

\affiliation{$^{3}$Institut Non-Lin\'eaire de Nice,
Universit\'e de Nice-Sophia Antipolis, 1361 Route des Lucioles, 06560 Valbonne, France}

\date{\today}

\begin{abstract}
We study Bragg spectroscopy of strongly interacting one dimensional
bosons loaded in an optical lattice plus an additional parabolic
potential. We calculate the dynamic structure factor by using Monte
Carlo simulations for the Bose-Hubbard Hamiltonian, exact
diagonalizations and the results of a recently introduced effective
fermionization (EF) model. We find that, due to the system's
inhomogeneity, the excitation spectrum exhibits a multi-branched
structure, whose origin is related to the presence of superfluid
regions with different  densities in the atomic
distribution. We thus suggest that Bragg spectroscopy in the linear
regime can be used as an experimental tool to unveil the shell
structure of alternating Mott insulator and superfluid phases
characteristic of trapped bosons.
\end{abstract}
\maketitle

\section{Introduction}
Cold atoms in optical lattices provide a way for realizing
interacting many body systems in essentially defect free
lattices~\cite{Jaksch}.  Paradigms of strongly correlated phenomena
such as the superfluid (SF) to Mott insulator (MI) quantum phase
transition have been realized in a three-dimensional lattice thanks
to the successful application of atom optics techniques to
traditional condensed matter systems~\cite{Greiner}. Recently, much
interest has been generated by experiments in reduced dimensionality
\cite{Tolra,Paredes,Weiss,Moritz,Fertig,Stoeferle}, where the role
of interactions and quantum fluctuations is enhanced. Of particular
relevance have been the realization of a SF/MI quantum phase
transition in one-dimension (1D) \cite{Stoeferle,Paredes}, and the
observation of a gas of hard core bosons \cite{Tolra,Paredes,Weiss},
or Tonks-Girardeau (TG) gas \cite{Girardeau}. The latter is a
characteristic of strongly interacting one dimensional systems,
 where the large repulsion  between atoms  mimics the
Pauli exclusion principle.
As a
consequence, there is a one to one correspondence of the
eigenenergies and eigenfunctions of TG bosons and non-interacting
fermions. This correspondence holds for all local observables, such
as the atomic density and fluctuations \cite{Girardeau}.

The presence in experiments of a confining parabolic potential
superimposed on the lattice provides the possibility of studying
strongly correlated systems at arbitrary densities. In fact, the
presence of the quadratic potential typically induces the
coexistence of alternating superfluid and insulating phases with
on-site lattice densities which can be larger than one
\cite{Jaksch,Kashurnikov,Batrouni,Demarco}. This creates a shell
structure, which is reminiscent of the MI lobes of the homogeneous
phase diagram \cite{Fisher}. In this case, the SF/MI transition is
better understood as a crossover than as a phase
transition~\cite{Batrouni}. Recently, it was shown that a system of
strongly interacting 1D bosons at high densities has the structure
of an array of stacked disjoint TG gases \cite{Guido2005}.  This
decomposition of the system in independent TG gases, or {\it
effective fermionization} (EF), allows for the computation of static
properties such as the density profile and fluctuations, as well as
dynamical properties.  In particular, EF has been used to explain
the microscopic mechanisms responsible for the decay of the
superfluid current of an interacting bosonic gas in a periodic
potential~\cite{Fertig}.

It remains a challenge to engineer experimental probes for atoms in
the strongly correlated regime.  Information on the excitation
spectrum has been obtained by using Bragg spectroscopy
\cite{Stoeferle}, sparking considerable theoretical activity
\cite{StoofBragg,AnaBragg,BatrouniBragg,IucciBragg}.  Unfortunately,
the experiment of Ref.~\cite{Stoeferle} was conducted in a regime
far from linear, which prevented the direct comparison of the
theoretical results with the experimental data. In particular, in
the case of the inhomogeneous system it has not been possible to
verify the nature of the double peak structure found in
Ref.~\cite{BatrouniBragg} in the low frequency response, by direct
comparison to the existing data of Ref.~\cite{Stoeferle}.

In this paper we study Bragg spectroscopy of bosons in the periodic
plus quadratic potentials in the linear response regime. We perform
quantum Monte Carlo simulations and exact diagonalizations of the
Bose-Hubbard Hamiltonian in the strongly interacting regime to
compute the dynamic structure factor. The latter is found to exhibit
a multi-branched structure  at small excitation frequencies
\cite{BatrouniBragg}. We  compare these numerical results  to the
predictions of the EF model, finding good agreement. The central
result of this paper is that this agreement strongly indicates that
this multi-branched structure at low-frequency is dominated by the
excitations of the system's superfluid components. That is, the
observed different excitation branches can be attributed primarily
to the response of the superfluid components of the various layers
of the EF model, whose presence is directly linked to the existence
of a shell structure in the many body density profile. When there
are at least two superfluid regions with different mean particle
numbers, as is the case of Ref.~\cite{Stoeferle}, this result
differs substantially from previous results, where the presence of
the double-peak structure in the response to the Bragg perturbation
was linked to the presence of particle-hole excitations in the Mott
phase \cite{BatrouniBragg}. Therefore, because of the sensitivity to
the presence of superfluids with different particle numbers, we
suggest that Bragg spectroscopy in the linear regime is an ideal
tool to characterize the shell structure of atoms in the quadratic
potential (see also \cite{GerbierPRL,GerbierPRA,Sengupta}).
\\ 

The presentation of the results is organized as follows. In
Sec.~\ref{BH} we introduce the Bose-Hubbard Hamiltonian, shortly
review the physics of strongly interacting bosons in a lattice
 and explain the basic ideas of the effective
fermionization model. In Sec.~\ref{DSF} we introduce the dynamic
structure factor. In Sec.~\ref{HL} we compare the results of the EF
model and of exact diagonalizations for a small number of bosons in
a homogenous lattice. The various branches of excitation are
explained in terms of (low-frequency) excitations of the superfluid
and (higher-frequency) particle-hole excitations. The EF model is
shown to well reproduce the excitations of the superfluid. In
Sec.~\ref{IL} we compare the results of quantum Monte-Carlo
simulations for experimentally realistic systems to the results of
the EF model. When there are more than one superfluid regions with
different particle densities, we show strong evidence that the
low-frequency excitations are dominated by the excitations of the
various superfluids. We then propose an experiment in the linear
response regime which would detect the associated shell structure of
the many-body density profile. Finally, the conclusions are
presented in Sec.~\ref{CONCL}.

\section{Bose-Hubbard Hamiltonian} \label{BH}
The Bose-Hubbard ({\it BH})
Hamiltonian describes $N$ interacting bosons in a lattice potential
of $M$ sites \cite{Jaksch}
\begin{equation}
\hat{H}= \sum_{j} \left[ \Omega j^2 \hat{n}_j +
\frac{U}{2}\hat{n}_j\left( \hat{n}_j-1 \right)  -J  \left(
\hat{a}_j^{\dagger}\hat{a}_{j+1} +
\hat{a}_{j+1}^{\dagger}\hat{a}_{j}\right)\right].\\
\label{EQNBHH}
\end{equation}
Here $\hat{a}_j$ is the bosonic annihilation operator  at site $j$,
and $\hat{n}_j=\hat{a}_j^{\dagger}\hat{a}_j$. $\Omega$ is the
curvature of the confining potential and  $U$ and $J$ are  the
on-site interaction and hopping energies.

In the homogeneous system, $\Omega=0$, the effective strength of the
interactions depends on the atomic density $N/M$, and the ratio
$\gamma=U/J$. In particular, for $N<M$ the system is effectively a
strongly interacting TG gas if $U$ is larger than the lattice band
width $4 J$ \cite{Paredes}. Thus, $\gamma=\gamma_c \approx 4$
determines  the critical interaction strength  in low density
systems. In particular, the SF/MI transition occurs in a homogeneous
unit filled lattice at $\gamma \approx \gamma_c$. For arbitrarily
large densities, $n-1 < N/M \leq n$, with $n$ an integer larger than
one, the system enters the strongly correlated regime when $\gamma
\gg \gamma_c n$. In these cases standard fermionization techniques
are invalid, however the low energy physics can still be well
reproduced by considering the excess $M [N-(n-1)]$  bosons  as
non-interacting fermions with an effective kinetic energy $nJ$
sitting on a plateau formed by  $M(n-1)$ atoms frozen in a Mott
state with exactly $n-1$ atoms per site. The validity of this
approximation relies on the fact that  states with more than $n$
atoms per site are suppressed by a factor on the order of
$1/\gamma$. In other words, when $\gamma \gg \gamma_c n$,  the whole
many-body system can be conveniently visualized as composed of two
independent subsystems: the atoms frozen in the Mott state and the
extra TG bosons\cite{Guido2005}.

For most experiments, a parabolic magnetic confining potential of
frequency proportional to $\Omega$ is also present, so the density
profile varies across the lattice. In the trapped case the
conditions for the formation of a Mott state change dramatically.
For example, it is always possible to create a unit-filled MI if
$N<M$, by varying the depth of the lattice or the magnetic trap
frequency. In the trivial $J=0$ limit, the density profile becomes a
``cake'' structure with maximal occupation $n^*$ at the trap center.
In our model, we view the density as a "layer cake" of $n$ stacked
horizontal layers. When $J=0$, the atoms are completely frozen, and
each layer of the cake may be viewed as an independent Mott state
with $N_n$ unit-filled sites.

For moderate values of $J>0$, it has been shown theoretically that
the density profile still has a cake structure of coexisting
superfluid and  MI phases. In this case, the density profile can be
visualized as being composed of stacked horizontal layers, but
because atoms are no longer frozen, in general the layers are not
independent. However, if number fluctuations in adjacent horizontal
layers do not overlap in space, all layers can be treated
independently and, as in the homogeneous system, standard
fermionization techniques can be applied to each layer separately.
In this situation  single-particle solutions provide expressions for
all many-body observables.  We call this generalization of the
Bose-Fermi mapping {\it extended fermionization} (EF).

 We want to point out that at variance with the
original definition \cite{Guido2005}, and for the sake of
simplicity, we here refer to all kinds of strongly interacting gases
as EF, independently of the density and the presence or absence of
an external potential. In the low density limit $n=1$, EF reduces to
standard fermionization.

In the remainder of this paper we focus on this strongly correlated
regime, where $\gamma \gg \gamma_c n$. Because $J$ decreases
exponentially with increasing lattice depth, this regime is easily
attained experimentally.

\section{The dynamic structure factor} \label{DSF}
The dynamic structure factor at temperature $T$ is given by:
\begin{equation}
S(q,\omega)= \frac{1}{\mathcal{Z}}\sum_{ji}e^{-\beta E_i} | \langle
i | \hat{\rho}_q | j \rangle |^2 \delta(\hbar\omega-E_i+E_j),
\label{StructFact}
\end{equation} where, $\mathcal{Z}$ is the
canonical partition function, $\beta^{-1}= T k_B$, with $k_B$ the
Boltzmann constant and $T$ the temperature. Here, $E_j$ are the
eigenenergies of the many body eigenstates $| i \rangle$, and
 $\hat{\rho}_q = \frac{1}{M} \sum_{j} e^{i q d j}
\hat{a}^{\dagger}_j\hat{a}_{j}$ is the density fluctuation operator
with $\hbar q$ the quasi-momentum and $d$ the lattice constant. For
the quantum Monte Carlo simulations we use a World-line algorithm at
small finite temperature $Tk_B =0.1 J$
\cite{BatrouniBragg,BatrouniScalettar}, while for the exact
diagonalizations we use standard linear algebra techniques, and $T$
can be taken to be zero. Equation (\ref{StructFact}) describes the
response of the system to Bragg spectroscopy in the linear response
regime. It implies that the system responds  whenever the frequency
$\omega$ of the Bragg perturbation matches the energy difference
between two eigenstates.

For $\gamma \gg \gamma_c n$ and at low enough energy, eigenstates
are densely grouped in energy ranges of the order of a few $J$
\cite{Guido2003}, separated by an energy of the order of $U$.
Eigenstates in each energy range are linear combinations of Fock
states with the same number of empty sites, of singly occupied
sites, doubly occupied sites, and so on. We thus expect that in the
strongly correlated regime the system's response to the Bragg
perturbation mirrors the ``grouped'' structure of the many body
energy spectrum.

In standard fermionization, $S(q,\omega)$ is obtained by computing
Eq.~(\ref{StructFact}) for a system of non-interacting fermions
\cite{Vignolo}.

\begin{eqnarray}
S(q,\omega)&=& \sum_{nm}|\sum_j e^{-iqdj} \psi_j^{ (n)} \psi_j^{
(m)}|^2 f(E^{(n)}) [1-f(E^{m})]\times\notag
\\&&\delta(\hbar\omega-E^{(n)}+E^{(m)}). \label{StructFer}
\end{eqnarray}Here $\psi_j^{(n)}$  and  $E^{(n)}$ are the
 $n^{th}$ single-particle eigenmodes and eigenenergies
of Eq.(\ref{EQNBHH})  with hopping energies $J$ respectively and $j$
is the lattice site index. $f(E^{(n)})$ denotes the Fermi-Dirac
distribution.

For the cases when there are more than one atom per site, according
to the EF model the density profile can be visualized as being
composed of stacked horizontal layers. In the parameter regime where
number fluctuations in adjacent horizontal layers do not overlap in
space, all layers can be treated independently and the dynamical
structure factor of the overall system can be approximated by adding
the structure factors  of the $n$  different layers each one with
$N_n$ atoms. The expression for  $S(q,\omega)$ in a given $n$ layer
is exactly the same than Eq.(\ref{StructFer}) but replacing the
 eigenmodes and eigenenergies   by the single-particle  solutions  of
Eq.(\ref{EQNBHH}) with hopping energies $nJ$ . Also  the chemical
potential in the Fermi-Dirac distributions must be calculated  to
consistently have an average of $N_n$ atoms in each layer.

The independent  addition of the structure factor of each layer is
justified in the model, because EF takes into account only the
responses of the superfluids to the Bragg perturbation. Thus, when
the two superfluid regions are spatially well separated, matrix
elements coupling low-lying excitations of the two superfluids in
Eq.~(\ref{StructFact}) become zero, because of the vanishing overlap
of the wave functions.

\begin{figure}[b]
\begin{center}
\leavevmode {\includegraphics[width=3.5
in]{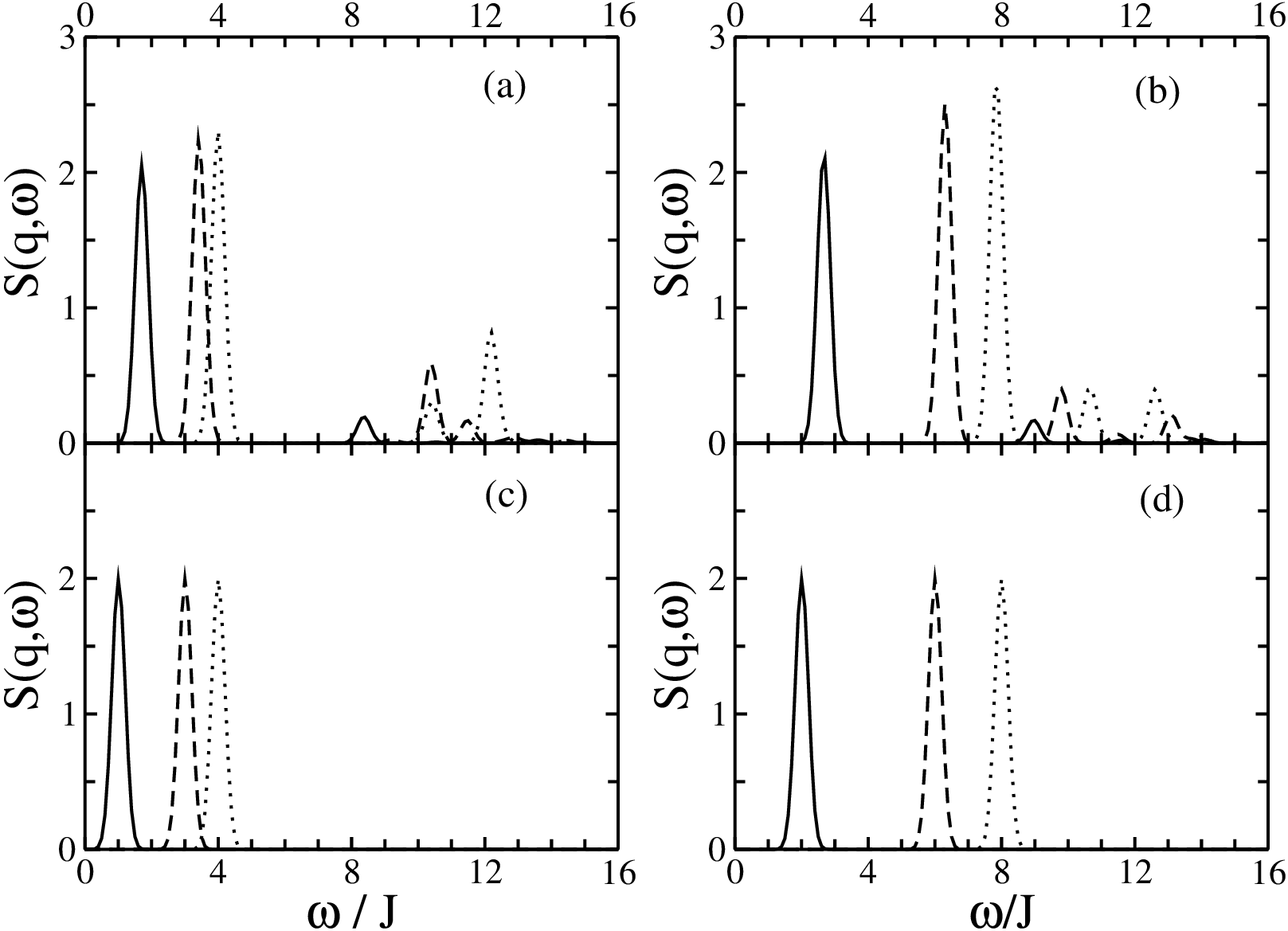}}
\end{center}
\caption{The dynamic structure factor $S(q,\omega)$ vs $\omega$,
with $qd =\pi/3$ (continuous line), $2\pi/3$ (dashed line) and
$\pi$ (dotted line). Here, $M=6$ in all panels, while $N=5$ (panels
(a) and (c)) and $N=7$ (panels (b) and (d)). Panels (a) and (b) are
results of exact diagonalizations with $\gamma=U/J=10$, while panels
(c) and (d) are the EF model results. For all plots, $\Omega/J=0$.
}\label{fig1}
\end{figure}

\section{Homogeneous lattice ($\Omega = 0$)} \label{HL}
In Fig.~\ref{fig1} we show $S(q,\omega)$ as a function of the Bragg
frequency $\omega$, for a homogeneous lattice with $M=6$ and
periodic boundary conditions. The number of atoms is $N=5$ (panels
(a) and (c)) and $N=7$ (panels (b) and (d)). The temperature $T$ is
equal to zero, in all the plots. The continuous, dashed and dotted
lines correspond to different quasi-momenta, with $qd = \pi/3, 2
\pi/3$ and $\pi$, respectively. Panels (a) and (b) are the results
of exact diagonalizations with $\gamma=10$, while panels (c) and (d)
are the predictions of the EF model for atoms with kinetic energy
$J$ and $2J$, respectively. In Fig.1(a) the exact results show large
peaks for $\hbar\omega/J \lesssim 4$, and smaller excitations for $8
\lesssim \hbar \omega/J \lesssim 14$. The value $4 J$ corresponds to
the single particle band width of a hole, that is of an empty site
tunneling in the lattice. These low frequency excitations are thus
due to the coupling between the ground state and eigenstates with at
most one atom per site. The smaller peaks in the interval $8
\lesssim \hbar\omega/J \lesssim 14$ are instead due to the coupling
to particle-hole excitations (1-ph), that is to eigenstates which
have one site occupied by two atoms and an extra empty site.
Accordingly, these peaks are centered around the value
$\hbar\omega=U$. The reduced  intensity of the 1-ph peaks  with
respect to the lowest energy peaks is expected, since in first-order
perturbation theory the coupling of the ground state to 1-ph is
proportional to $1/\gamma$. In fact as analytically calculated in
Ref.~\cite{AnaBragg},  for a unit filled system the height of the
particle hole peaks is proportional to $ {64/\gamma^2} $ which is of
the order observed in the plot. Panel (c) shows that the position
and the height of the lowest energy peaks are reasonably well
reproduced by the EF model, which here corresponds to standard
fermionization, since $n=1$. This is remarkable, since $\gamma$ is
only 10 in Fig.1(a), while it is considered infinite in the model.
On the other hand, the high energy peaks are missing, since 1-ph are
not present in the model \cite{Guido2005}.

The qualitative analysis above is easily adapted to the case $M=6$,
$N=7$ of panels (b) and (d). Here, $n=2$ and the EF model predicts
that the low energy spectrum is the one of a single particle with
hopping energy $2J$ on top of a plateau of 6 atoms, which are frozen
in a unit-filled Mott insulator. Thus, $S(q,\omega)$ in the EF model
reduces to computing Eq.~(\ref{StructFact}) for a non-interacting
fermion with hopping energy $2J$. Since the single particle band
width is now $8J$, we expect a large low frequency response in the
energy range $\hbar\omega/J\leq 8$. This is actually shown in
Figs.~\ref{fig1}(b) and (d), which confirms the validity of the EF
model even for such a small value of $\gamma$. In fact, the latter
is just barely larger than $2 \gamma_c=8$. In addition, the exact
results of Fig.~\ref{fig1}(b) show excitations around
$\hbar\omega\approx U$. Not surprisingly, the latter are due to
coupling to eigenstates with two sites occupied by two atoms, and to
eigenstates with a site occupied by three atoms. These couplings are
again proportional to $1/\gamma$, and thus the height of these high
energy peaks is highly suppressed. We have also found numerically
excitations at an energy of order $2 U$. Because the couplings of
these eigenstates to the ground state are of order $1/\gamma^2$,
these peaks are very small, and are not shown in the graphs.

Finally, we checked numerically that in the case $M=N=6$, where the
ground state is a Mott insulator with one atom per site, the lowest
energy peaks are centered around $U$. The spectrum is thus gapped,
in this case, as expected. The peaks' height is of the same order of
magnitude of the peaks of Figs.1(a) and (b) centered around
$\hbar\omega\approx U$. A comprehensive discussion of the
excitations in the MI phase can be found in Ref.~\cite{AnaBragg}. In
the following we show how the above  discussion  is useful in the
description of the spectrum of excitations when the quadratic trap
is present.

\section{Inhomogeneous lattice ($\Omega > 0$)} \label{IL}
\begin{figure}
\begin{center}
\leavevmode {\includegraphics[width=3. in]{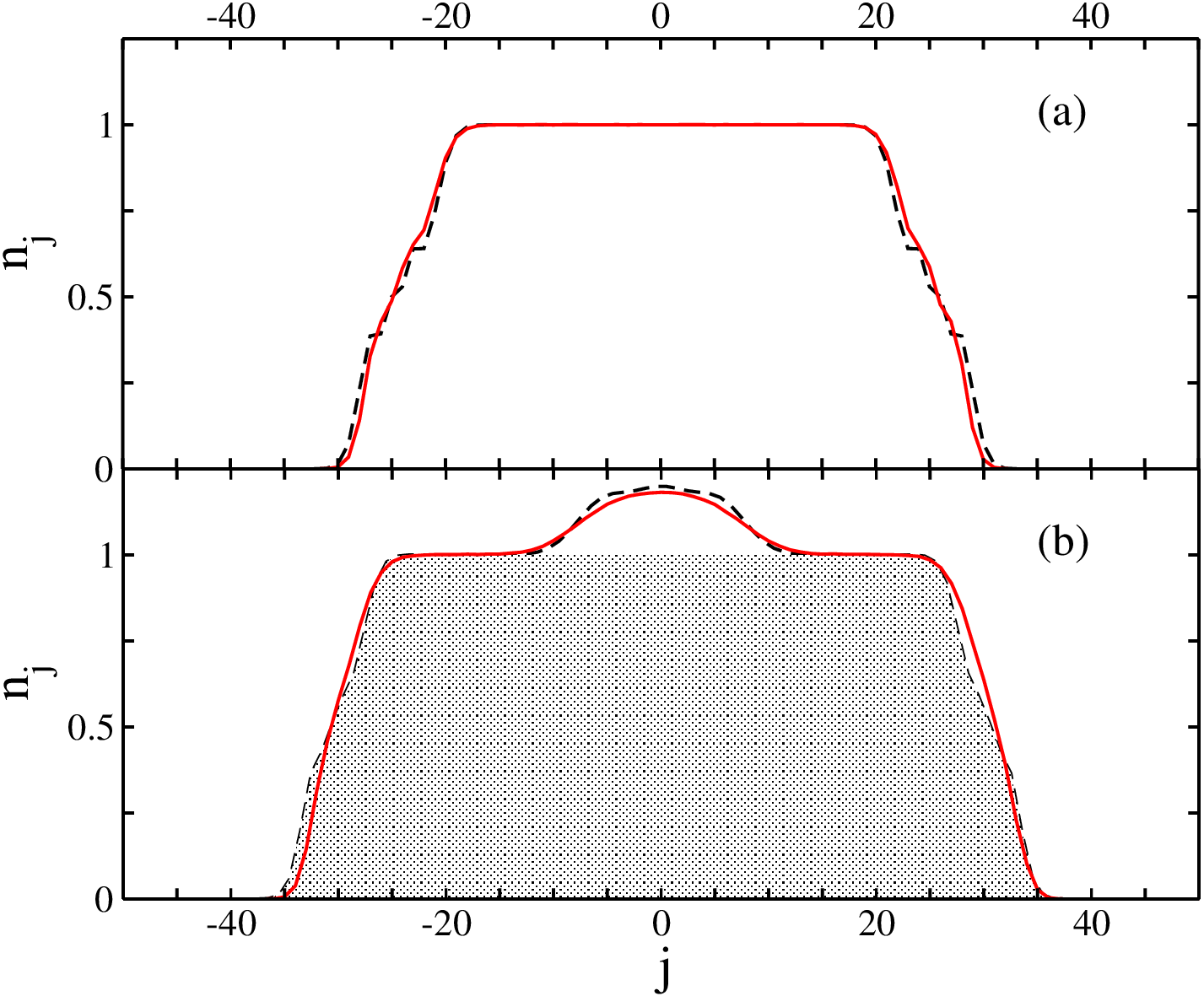}}
\end{center}
\caption{(Color online) Local density profile, $n_j$, as a function
of the lattice site $j$ for the confined system with $M=100$, and
$\Omega/J=0.008$. Panel (a): $N=50$ and $\gamma=14$. Panel (b):
$N=64$ and $\gamma=8$. The continuous (red) and dashed (black) lines
are the quantum Monte Carlo and EF model results, respectively.  The
shaded area in panel (b) corresponds to atoms in the lowest layer of
the EF model. These atoms are largely frozen in a unit-filled Motts
state, analogous to the case of panel (a).  }\label{fig2}
\end{figure}

When the quadratic trap is present and $\gamma > \gamma_c$, the
system can show the coexistence of superfluid and insulating phases.
Figure~\ref{fig2} displays the local density profiles of systems
with such coexisting SF and MI phases, with onsite peak density one,
(a), and larger than one, (b). In the plots, integer and non-integer
local densities signal MI and SF phases, respectively. Here, $M=100$
and $\Omega=0.008 J$. In Fig.~\ref{fig2}(a) the value of $\gamma$ is
14, while it is $\gamma=8$ in Fig.~\ref{fig2}(b). These values and
the fillings we use are comparable to those used in current
experiments. The continuous (red) and dashed (black) lines are the
quantum Monte Carlo data and the predictions of the EF model,
respectively.  The two curves are almost indistinguishable on the
scale of the graph, which is remarkable at least for the results of
panel (b), where the EF model at the center of the trap is barely
applicable.  For the case of Fig.~\ref{fig2}(a) where the peak
on-site density is one, the EF model corresponds to standard
fermionization. On the other hand, for the case of
Fig.~\ref{fig2}(b), the EF model amounts to decomposing the density
profile into two vertically stacked horizontal layers of atoms
\cite{Guido2005} (with $N_1=61$ and $N_2=3$ atoms in the lower and
upper layers, respectively).  In Fig.~\ref{fig2}(b) atoms in the
lowest layer have been shaded, in order to visualize the two layers
of the EF model.  In each layer there is at most one atom per site,
and thus the EF idea is to apply standard fermionization techniques
to each layer separately.
\begin{figure}
\begin{center}
\leavevmode {\includegraphics[width=3. in]{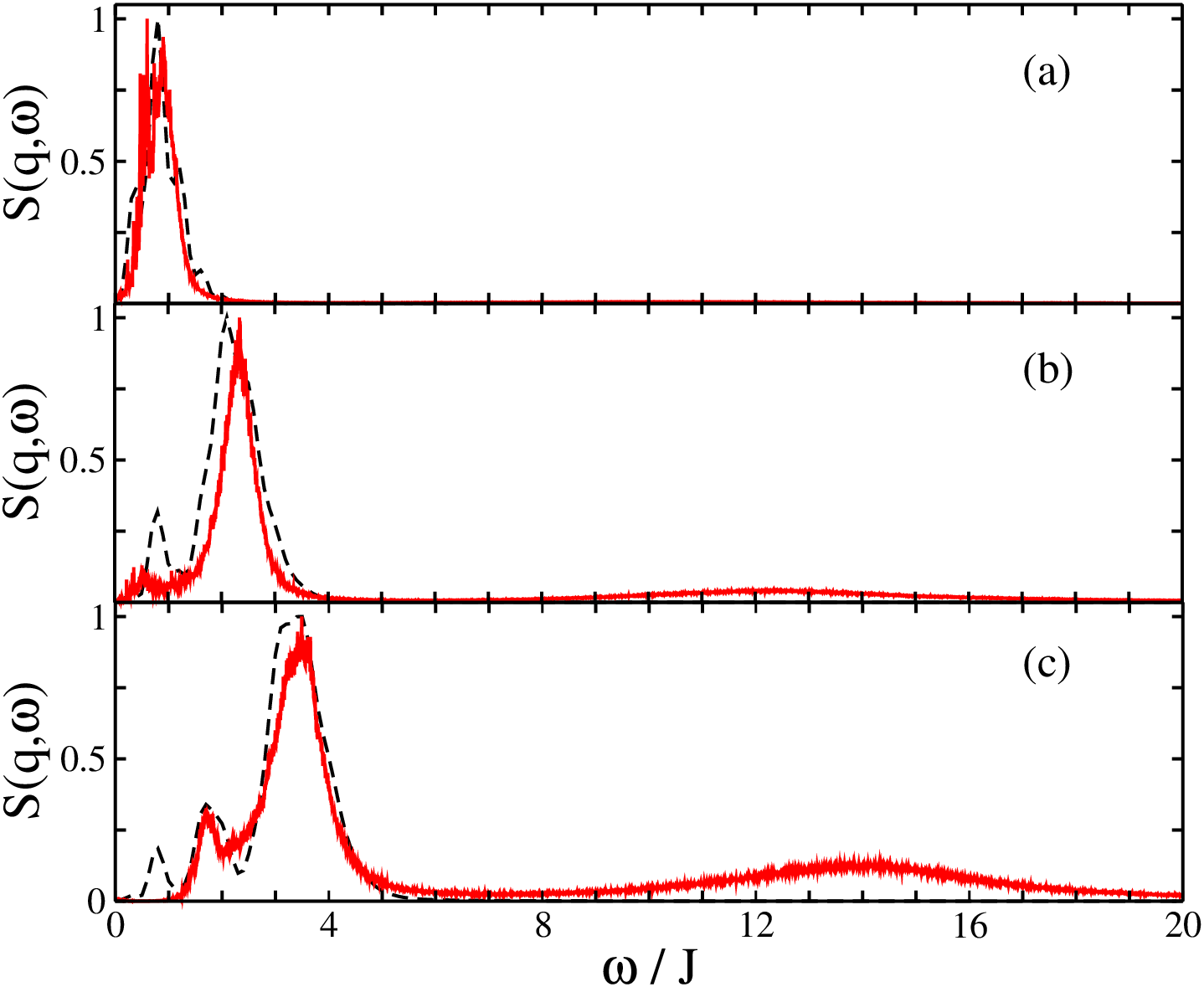}}
\end{center} \caption{(Color online) $S(q,\omega)$ vs $\omega$ for
the system in Fig.~\ref{fig2}(a) with $N=50$, $\gamma=14$, and for
$qd=0.2 \pi$(a), $0.5 \pi$(b), and $0.9 \pi$(c). The countinous(red)
and dashed(black) lines are the quantum Monte Carlo and EF model
results, respectively. The large excitation peaks for
$\hbar\omega/J<4$ are excitations of the superfluid component, while
the broad peak at $\hbar\omega\approx U=14 J$ is due to
particle-hole excitations of the combined SF and MI, analogous to
Fig.1.
 }\label{fig3}
\end{figure}
Then, it is possible to calculate many body observables (e.g., the
density profile) for the two layers separately, and add the results
to obtain the total value. Details of the computation for the
density profile are given in Ref.~\cite{Guido2005}. There, it is
shown that EF provides accurate results whenever the superfluid
regions at the trap center and at the edges are spatially well
separated by sufficiently large MI regions. In fact, in this case it
is possible to unanbiguously attribute a certain number of atoms to
each layer of the EF model, so that layers are well defined. This is
the case of Fig.~\ref{fig2}(b), where an insulating region of about
10 lattice sites separates the superfluids at the trap center and at
the edges. In the following we compute $S(q,\omega)$ for the two
layers separately and then add the obtained results to calculate the
complete $S(q,\omega)$.

Figure \ref{fig3} shows $S(q,\omega)$ as a function of $\omega$ for
some values of $q$ for the system of Fig.2(a), where there is at most
one atom per site. In particular, Figs.~\ref{fig3}(a), (b) and (c)
correspond to $qd=0.2\pi,0.5 \pi$ and $0.9\pi$, respectively. In the
calculations we have normalized the height of the largest peak to
1. Analogous to the case of Fig.~\ref{fig1}(a), the numerical solution
shows large peaks in the energy range $\hbar \omega \leq 4 J$. The EF
solution, which here reduces to standard fermionization, captures the
location and width of these peaks rather well. Since fermionization
accounts only for couplings to states with at most one atom per site,
we conclude that the response for $\hbar \omega/J \leq 4$ is solely
due to the small number of atoms in sites $|j| \geq 20$ which are in
the superfluid phase (analogous to the discussion of
Fig.~\ref{fig1}(a)).  These are thus ``surface'' excitations.  In
addition, for $q$ large enough ($qd\sim \pi$) the numerical solution
shows the presence of peaks around $U$, which are not captured by the
EF model. These are due to couplings to 1-ph excitations, in analogy
to the discussion above. Therefore, the excitation spectrum
exhibits two branches, a lower energy one corresponding to excitations
of the surface superfluid layer and a higher one corresponding to the Mott
region.

The differences between the numerical and EF results can be
attributed both to errors of order $\sim 1/\gamma^2$ in the EF
predictions and to the fact that the numerical calculation of
$S(q,\omega)$ is difficult for $\gamma \gg 1$. In this respect, we
notice that the lowest energy excitation in the EF model is always
at $\hbar \omega\approx \Omega N=0.4 J$, which is the energy cost
for an atom at position $N/2$ (farthest outlying occupied site), to
tunnel to the next unoccupied site. The fact that the numerical
results do not grab this excitation for $qd \sim \pi$
(Fig.~\ref{fig3}(c)) suggests a partial failure of the numerics.

\begin{figure}
\begin{center}
\leavevmode {\includegraphics[width=2.25 in]{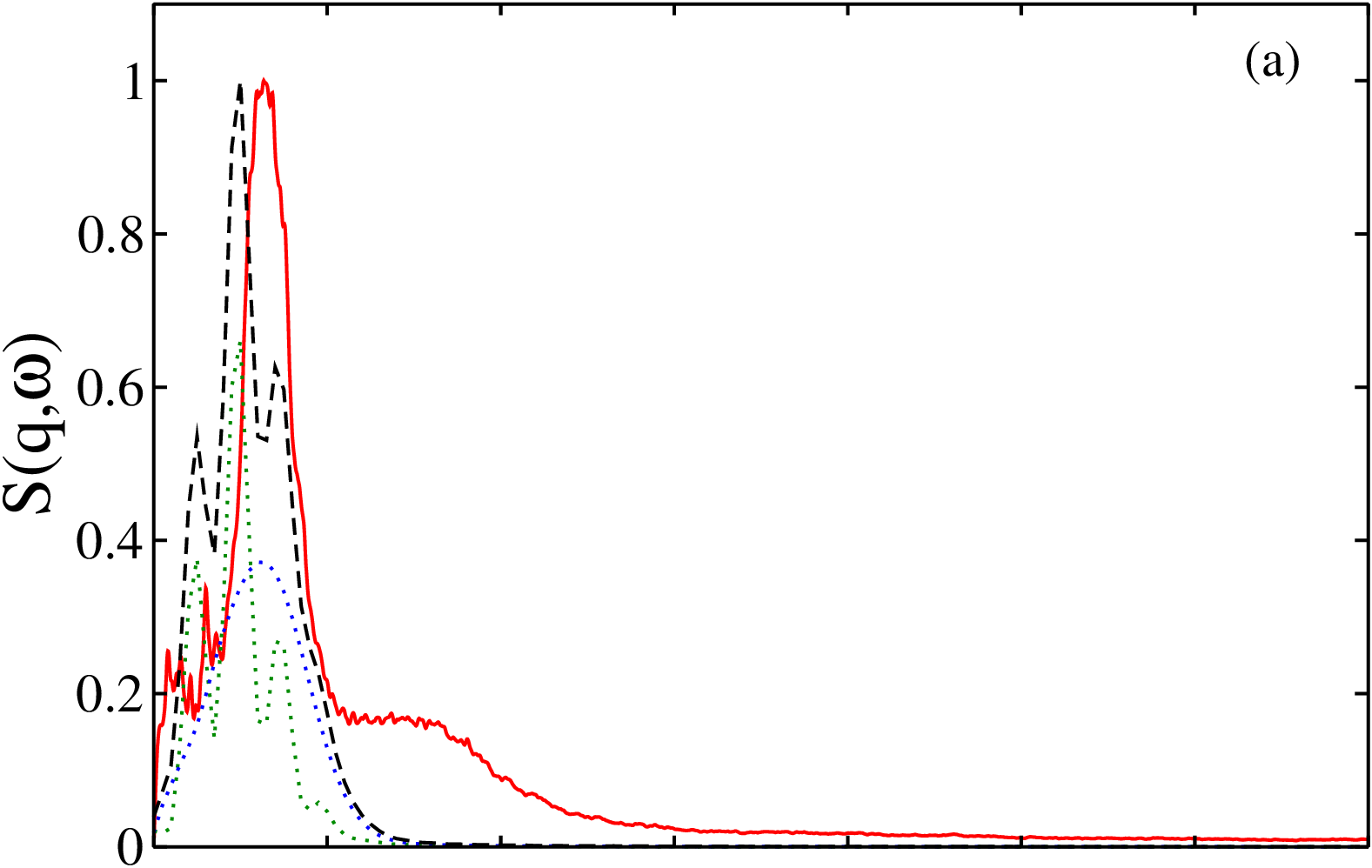}}
\end{center}

\begin{center}
\leavevmode {\includegraphics[width=2.25 in]{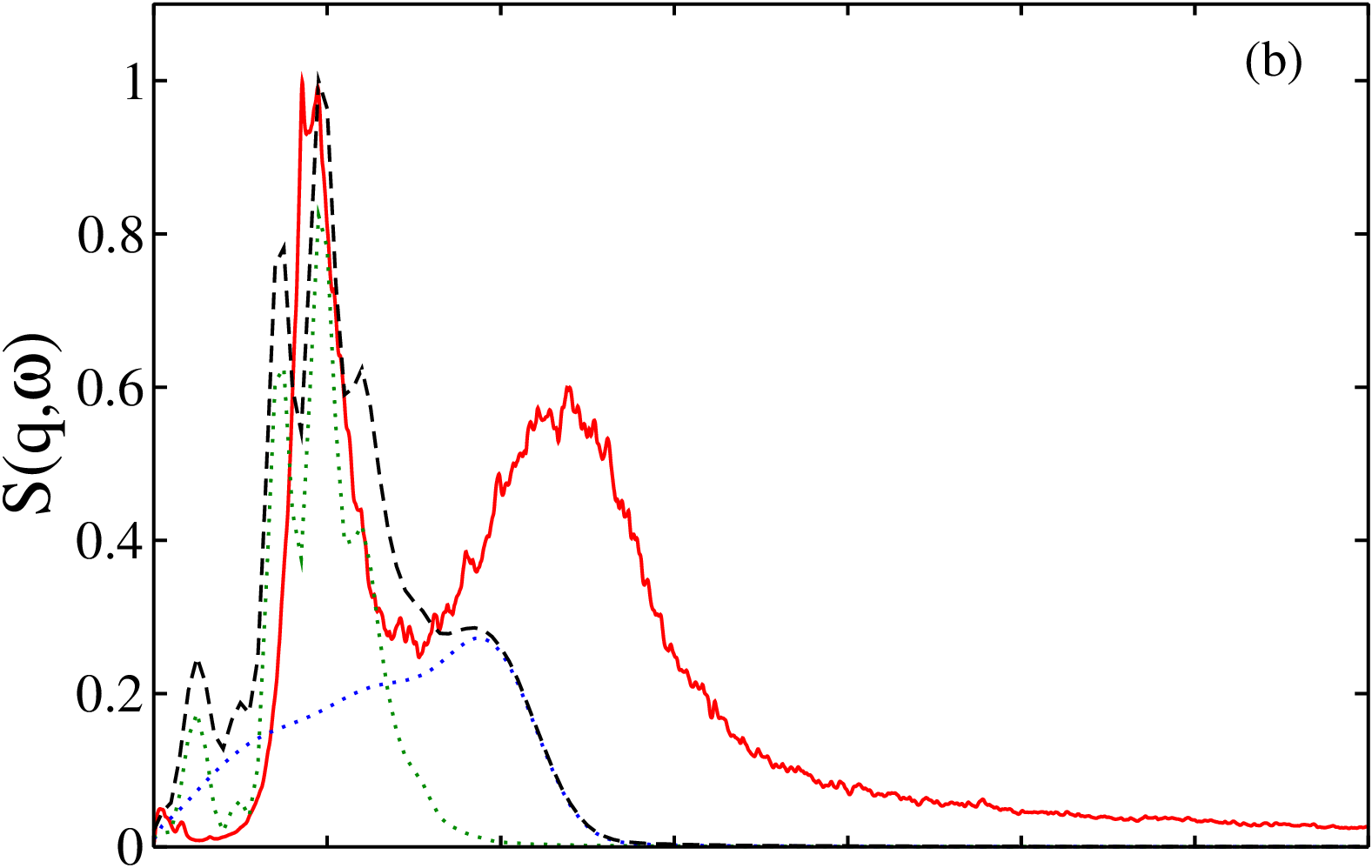}}
\end{center}

\begin{center}
\leavevmode {\includegraphics[width=2.25 in]{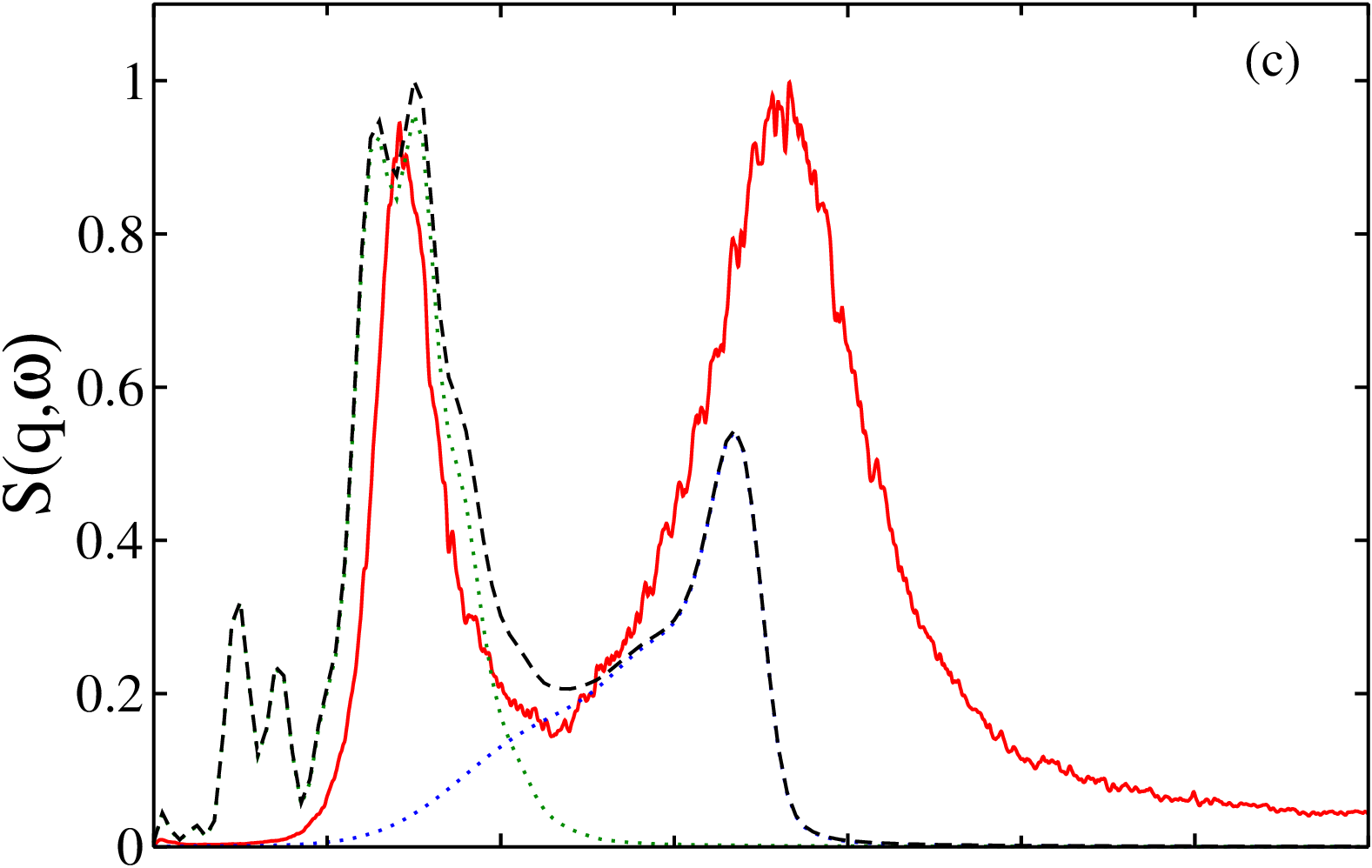}}
\end{center}

\begin{center}
\leavevmode {\includegraphics[width=2.25 in]{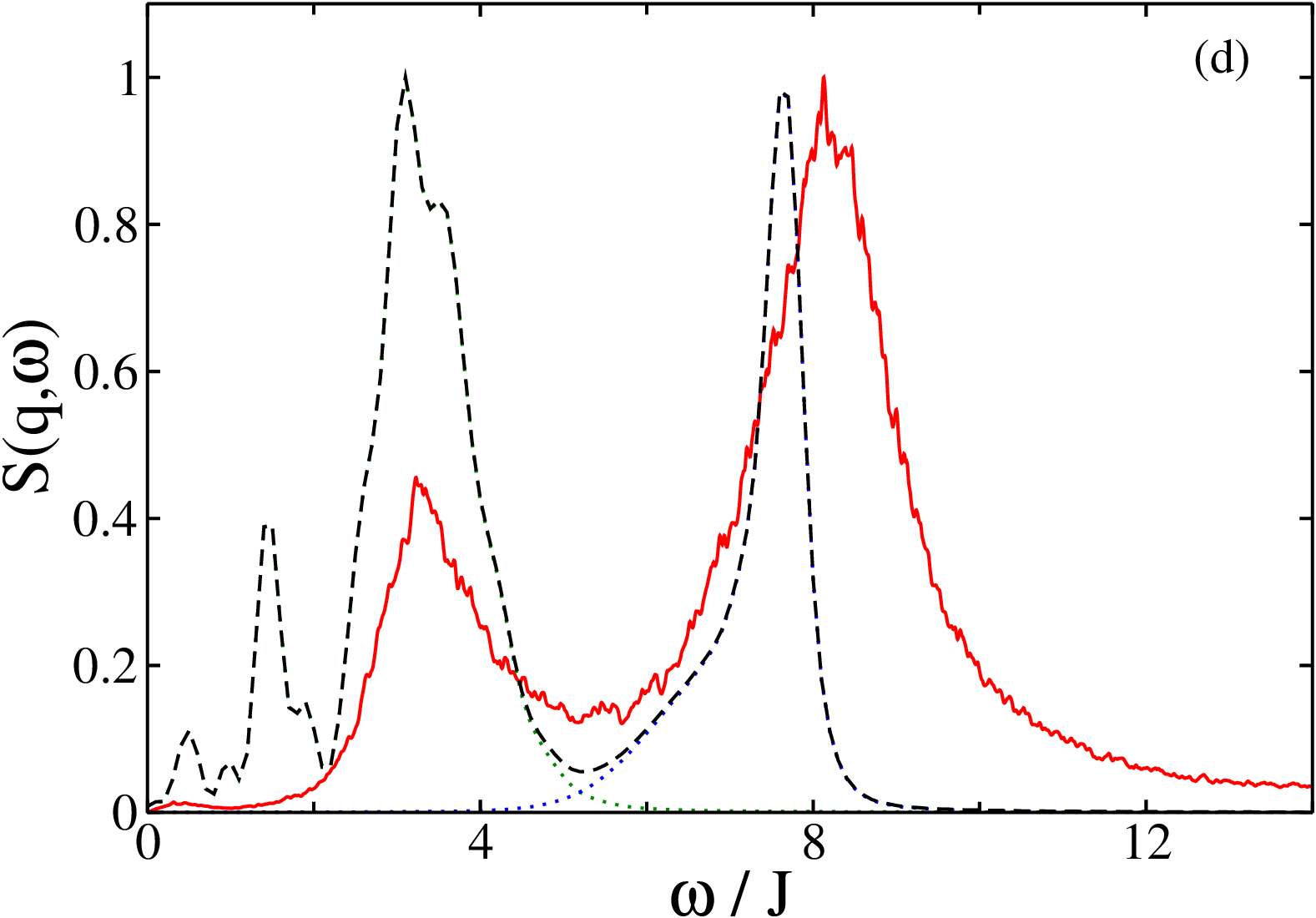}}
\end{center}
\caption{(Color online) $S(q,\omega)$ vs $\omega$ for the system in
Fig.~\ref{fig2}(b) with $N=64$, $\gamma=8$, and for $qd= 0.2\pi$(a),
$0.4 \pi$(b), $0.7 \pi$(c), and $0.9 \pi$(d). The countinous (red)
and dashed (black) lines are the quantum Monte Carlo and EF model
results, respectively. The dotted lines are results for the lower
(green) and upper (blue) layers of the EF model. The dashed line is
thus obtained by adding these curves. The excitation peak saturating
at $\hbar\omega/J\approx 4 $ is due to excitations in the SF with
less than one atom per site, ($n_j <1$, with $25 \leq |j| \leq 35$
in Fig.~\ref{fig2}(b)). The excitation peak saturating at
$\hbar\omega/J\approx 8 $ is due to excitations in the SF with more
than one atom per site, ($n_j > 1$, with $0 \leq |j| \leq 10$ in
Fig.~\ref{fig2}(b))). Because $\gamma=8$, particle-hole excitations
around $\hbar\omega/J \approx 8$ increase the width of this SF peak
in the Monte Carlo results. }\label{fig4}
\end{figure}
Similarly, Fig.~\ref{fig4} shows $S(q,\omega)$ (normalized to 1) as a
function of $\omega$ at different values of $q$ for the case of
Fig.~\ref{fig2}(b), where there is a superfluid phase at the trap
center with more than one atom per site. The continuous (red) line is
the Monte Carlo result, while the dashed (black) line is the EF model
result. As mentioned above, the EF model assumes that the system of
Fig.~\ref{fig2}(b) is conveniently decomposed into two disjoint,
vertically stacked gases, each one with at most one atom per
site. Most atoms in the lowest layer are frozen in a Mott insulator
state, in analogy to the case of Fig~\ref{fig3}(a), while atoms in the
upper layer are delocalized in a region approximately comprising the
sites $-10 \leq j \leq 10$. Atoms in the lower and upper layers have
hopping energies $J$ and $2J$, respectively, and are treated as
independent. The results for $S(q,\omega)$ for the lower and upper
layers are indicated by dotted lines (green and blue lines
respectively).

Figure~\ref{fig4} shows that the numerical excitation spectrum
(continuous line) at large enough momenta has two branches of
excitations, which saturate at $\hbar \omega\approx 4J$ and $8 J$,
respectively. The existence of these two branches of excitation has
been previously reported by one of us in Ref.~\cite{BatrouniBragg},
where the lower and higher frequency peaks were attributed to
excitations of the SF and MI phases respectively, as in the case of
Fig.~\ref{fig3}. However, while this interpretation is correct for
Fig.~\ref{fig3}, it is only partially so in this case. The good
agreement between the numerical results and the results of the EF
model (dashed line) with respect to the position and width of the
large feature around $\hbar \omega\approx 4J$, and the position of
the peak around $\hbar \omega\approx 8J$ for large $q$,  strongly
suggests that this double peak structure is mainly due to
excitations of the superfluid components of the two layers of the EF
model. In particular, the lower frequency excitations ($\hbar \omega
\leq 4 J$) are due to superfluid atoms in the lowest layer of the EF
model (sites where $n_j<1$, in the shaded region of
Fig.~\ref{fig2}), while the higher frequency excitations are mostly
due to excitations of atoms in the upper layer, which are all in the
SF phase. As mentioned above, the saturation thresholds $4J$ and
$8J$ correspond to the values of the effective single particle
band-widths of atoms with hopping energy $J$ and $2J$, respectively.

Since here $\gamma=8$, 1-ph excitations certainly do contribute to
the system's response around $\hbar \omega/J=8$.  However, since the
coupling to 1-ph excitations in the linear regime is rather
suppressed, as can be seen in Fig.~\ref{fig3} (it is proportional to
$1/\gamma$), we argue that the intensity of the response peak
centered around $\hbar\omega \approx 8J$ is dominated by the
excitations of SF atoms in the second layer rather than by 1-ph
excitations. Unfortunately, a detailed discussion of Fig.~\ref{fig4}
is complicated by the fact that the discrepancies between the model
and Monte-Carlo results are most likely of physical as well as
computational nature. As said above, the physical origin of the
discrepancies resides in the EF model neglecting 1-ph excitations.
This contributes to the underestimation of the height and width of
the peak around $\hbar \omega \approx 8J$ in the EF results of
Fig.~\ref{fig4}(c) (while the peak around $\hbar \omega \approx 4 J$
is well reproduced). On the other hand, because the relative height
of the peaks in the numerical solution varies greatly from $q d =0.7
\pi$ (panel (c)) to $q d =0.9 \pi$ (panel (d)) and oscillates in
between (not shown), we are also led to suppose the existence of
some instability in the numerical results. This prevents us from
attributing unambiguously all the discrepancies between the
numerical and model results to 1-ph excitations, and thus getting a
quantitative estimate of 1-ph contributions to the excitation
spectrum.

With all the $provisos$ above, we can safely state that an important
effect of coupling to 1-ph excitations is to increase the width of
the peak around $\hbar \omega \approx 8J$ with respect to the EF
results. That is, by increasing the ratio $\gamma$, we would expect
to see a continuous decrease of the width of the peak at
$\hbar\omega/J=8$, and the growth of a distinct peak at the chosen
$\gamma$-value. This new distinct peak would be due to 1-ph
excitations of the atoms in the two layers, analogous to the
discussion of Fig.~\ref{fig1}(a-b).

In other words, the presence of a multiply peaked structure in the
low frequency response  to the Bragg perturbation for $\gamma \gg
nJ$ is clear evidence of the existence of a shell structure in the
many body density profile (Fig.~\ref{fig2}). In fact, for $qd
\approx \pi$ the peak corresponding to the superfluid component of a
generic layer $n$ saturates at $ \hbar \omega \approx 4nJ$, and thus
the presence of low-frequency peaks separated by $4J$ is an
indication of the cloud's shell structure. We thus propose to use
Bragg spectroscopy to characterize the various phases of the trapped
strongly correlated bosons, by detecting this multiple peak
structure. For example, the system's response could be measured in
successive experiments while keeping fixed the trapping potentials
(and thus the $\gamma$ and $\Omega/J$ ratios), and varying the total
number of atoms $N$. Then, for {$q d\approx \pi$ and
$N<\sqrt{4U/\Omega}$ a single peak at $\omega\approx 4J$ would be
observed ($\sqrt{4U/\Omega}$ is a threshold value for having only
one layer in the EF model, \cite{Guido2005}). By increasing $N$ to
values larger than $\sqrt{4U/\Omega}$ a peak at
$\hbar\omega/J\approx8$ would be suddenly observed in the system's
response, signaling the formation of a SF with two atoms per site at
the trap center. By further increasing $N$, other peaks would occur
at multiples of $4J$, signaling the formation of superfluid regions
with more than two atoms per site. It should be noted that typically
the lifetime of these states would be severely limited by three-body
recombination, and the observation of these layers may become
complicated.

Finally, we notice that Bragg spectroscopy in the linear regime may
also be used to detect the atoms' shell structure in dimensions $D$
larger than one (where fermionization techniques are not
 applicable). In fact, in the appropriate parameter regimes,
the system retains the structure of an array of vertically stacked
horizontal layers. Thus, the response to the Bragg perturbation of
atoms in the superfluid regions of the different layers should
provide for a signature of the shell structure. Since the kinetic
energy is a factor of $D$ larger, the various branches of
excitations at low frequency should be more separated for $D>1$ than
for $D=1$. This may facilitate the experimental resolution of the
excitation peaks in dimensions higher than one.

\section{Summary} \label{CONCL}
We have presented a study of the dynamical
structure factor for strongly interacting bosons confined in
homogeneous lattices and in presence of a quadratic confining
potential. By comparing the results of quantum Monte Carlo
simulations and exact diagonalizations with the predictions of an
analytical EF model, we characterized the excitation spectrum of
strongly correlated bosons at arbitrary densities. In particular, we
were able to identify low frequency peaks in the system's response
to the Bragg perturbation which are clear cut evidence of the
formation of a shell structure in the density profile. This shell
structure is made of alternating SF and MI phases, which we propose
to study
using Bragg spectroscopy in the linear response regime.\\

We thank R. T. Scalettar, F. F. Assaad and M. K\"ohl for very
helpful discussions. G.P. acknowledges support from the EU through a
grant of the OLAQUI Project. A.M.R. acknowledges support from a
grant of the Institute of Theoretical, Atomic, Molecular and Optical
Physics at Harvard University and Smithsonian Astrophysical
observatory.

\end{document}